\documentclass[aps,twocolumn]{revtex4}
\usepackage{amsmath,amssymb}

\usepackage{epsfig}
\usepackage{graphicx}

\def\be{\begin{equation}}
\def\ee{\end{equation}}
\def\bea{\begin{eqnarray}}
\def\eea{\end{eqnarray}}

\newcommand\fverb{\setbox\pippobox=\hbox\bgroup\verb}
\newcommand\fverbdo{\egroup\medskip\noindent                        \fbox{\unhbox\pippobox}\ }
\newcommand\fverbit{\egroup\item[\fbox{\unhbox\pippobox}]}

\newcommand{\bear}{\begin{eqnarray}}
\newcommand{\eear}{\end{eqnarray}}

\newbox\pippobox

\def\6{\partial}

\def\a{\alpha}
\def\nn{\nonumber}

\def\sq
\def\a{\alpha}

\begin{document}

\title{A solution of the coincidence problem based on the recent galactic
core black hole mass density increase}

\author{Georgios Kofinas}\email{gkofin@phys.uoa.gr}
\affiliation{Department of Physics, University of Crete, 71003
Heraklion, Greece}

\author{Vasilios Zarikas}\email{vzarikas@teilam.gr}
\affiliation{Department of Electrical Engineering, ATEI Lamias,
35100 Lamia, Greece}

\date{\today}




\begin{abstract}

A mechanism capable to provide a natural solution to two
major cosmological problems, i.e. the cosmic acceleration and the coincidence problem, is proposed.
A specific brane-bulk energy exchange mechanism produces a total dark pressure, arising when adding all
normal to the brane negative pressures in the interior of galactic core black holes. This astrophysically
produced negative dark pressure explains cosmic acceleration and why the dark energy today is of the same
order to the matter density for a wide range of the involved parameters.
An exciting result of the analysis is that the recent rise of the galactic core black hole mass density
causes the recent passage from cosmic deceleration to acceleration. Finally, it is worth mentioning that
this work corrects a wide spread fallacy among brane cosmologists, i.e. that escaping gravitons result to positive
dark pressure.

\end{abstract}


\maketitle

\section{Introduction}

During last decades it has been realized that the investigation of
the problems associated with the cosmological constant would provide
an insight into the structure and the properties of elusive quantum
gravity. A serious problem concerning the cosmological constant
refers to the vast discrepancy between the value a
theorist would expect and the very low value of the effective
cosmological constant. As Zel'dovich \cite{zeld} first noticed, the
effective cosmological constant we measure is the sum of the pure
geometric origin cosmological constant plus the energy density of
the vacuum. It seems impossible to understand why the measured
effective cosmological constant is so much smaller than the value of
the vacuum energy calculated by a quantum field theorist (cosmic
phase transition, quantum field zero-point energies). This puzzle
challenges the inflationary scenario and the various models of
quantum gravity.

The present work attempts to solve a recently emerged problem
regarding the cosmological constant issue which concerns the measured
cosmic acceleration. During the last decade, it has been established through different
independent pieces of astronomical data that empty space, devoid of
the usual matter, is anti-gravitating. It creates gravitational
repulsion and gives rise to an accelerated cosmological expansion.
According to our present-day understanding, this accelerated
expansion could be induced by an effective cosmological
``constant"-like term (vacuum $p=-\rho $ or dark energy $p<-\frac{\rho }{3}$).
Data suggest that the magnitude of the required vacuum/dark energy is quite
close to the critical (closure) cosmological energy density (approximately 70\%).
Why vacuum energy, which stays
constant in the course of cosmological evolution, or why dark
energy, which evolves with time quite differently from the normal
matter, have similar magnitude with matter density just today, all
being close to the value of the critical energy density?

There are several ideas in literature, though yet incomplete, that
have the potential to provide solutions to cosmic acceleration problem. The present paper proposes that
\emph{the negative five-dimensional pressure produced from an astrophysical brane-bulk energy outflow
occurring inside all cosmic black holes} is large enough to drive the measured cosmic acceleration.
Assuming an RS-like cosmological brane \cite{RS}, \cite{Roy}, \cite{gregory}, this total pressure
(called dark pressure) arising from the sum of all negative pressures normal to the brane suffices
to explain the coincidence problem, without using
a negative vacuum energy or exotic fields/fluids throughout the universe. In the presented scenario
dark energy ``originates" from dark matter (grows from a negligible value to a significant one due to dark pressure); moreover, the recent appearance of the cosmic acceleration
is correlated to the recent increase of the galactic core black hole density.
An accelerating universe has already been produced
by several brane-bulk energy exchange scenarios \cite{zarikas}, \cite{tamva}, \cite{tetradis},
\cite{langlois}, \cite{Hebecker}. However, in all these works the whole universe should be hot enough,
and therefore, these scenarios fail to explain recent acceleration.

Both astrophysical black holes in haloes and supermassive black
holes at the galactic centres appear after the large scale structure
of the universe, weight a portion of $\rho _{m}$ and are regions
where high energy interactions occur. This fact will be at the
center of the proposed mechanism. Assuming that a brane cosmological
model describes our universe, it is natural to expect a moderate
exchange of energy between the brane and the bulk. Astrophysical black holes
contain matter in an unknown form, i.e. effective quantum fluid (possibly
arising from superposition of non empty black hole quantum
spacetimes) and accrete continuously mass. Collapsing matter falling
into a black hole accelerates, interacts and gets easily ``thermalized" to
temperatures close and above $M$ (five-dimensional fundamental Planck mass).
Furthermore, it is expected portion of black hole mass to be in the
form of highly energetic states close to $M$, not only due to
accreting matter interactions but also due to Hawking-like particle
production in the interior \cite{greenwood}. But for energy scales
close to $M$, matter interactions result to graviton escape
to the bulk. Therefore, energy outflow can occur in the interior of
galactic halo black holes and galactic core supermassive black
holes. This black hole originated exchange results to non-zero energy-momentum tensor components
$T_{05}$ and $T_{55}$ and is able to provide the necessary conditions for a cosmic acceleration.

The emission of gravitons to the bulk is associated with negative dark pressure on the brane.
Indeed, a brane experiences positive pressure when bulk particles fall into it. This negative pressure
is the responsible quantity that provides the required amount of the measured cosmic acceleration.
Although there is a small outflow in our scenario in each of the galactic black holes (consistent with
black hole mass evolution and galaxy dynamics), this leakage is associated unavoidably with orders of
magnitude larger dark pressure. Moreover, the emerged dark energy is not a small portion of the dark matter but of the same order with it. Note that in our scenario, before outflow starts
(before galaxy formation) there may be either a zero or a very small positive non zero decelerating dark radiation term $\frac{\mathcal{C}}{a^4}>0$. However, this radiation term overpasses well known problems of nucleosynthesis constraints \cite{ichicki}.
When negative dark pressure emerges, this radiation term is modified and finally becomes an accelerating dark energy component.

Note also that it costs almost nothing to stretch a brane that has zero total tension/cosmological constant
\cite{thooft}. Apparently spacetime is such that it takes a lot of energy to curve it, while stretching it is
almost for free, since the cosmological constant is zero. This is quite contrary properties of objects from
every day experience, where bending requires much less energy than stretching. The present study tries only to explain
naturally the recent cosmic acceleration while assumes that there is some mechanism that sets the cosmological constant from field theory vacuum energy equal to zero (for example R-S fine tuning, holography etc.).

The paper is organized as follows. In section II the
mathematical framework is presented. Here, it becomes obvious from Eq. (\ref{simple2}) that the negative dark
pressure $\Pi$ can drive acceleration. In addition, Eq. (\ref{systemb3}) shows that $\Pi$ determines also the
time derivative of dark energy, resulting to the current value of $w_{DE}$. Furthermore, the question of how
large should be the present value of dark pressure $\Pi$, is estimated in Eq. (\ref{current po}). Such a value
can easily be obtained, as it is explained in sections III and V, for moderate values of the involved astrophysical
parameters. In section III the connection of the mathematical framework with the astrophysical context is given.
Section IV provides various supportive theoretical aspects of the physics of the proposed mechanism. However,
section IV is not crucial to the main results of the analysis.  In section V numerical computations are presented
in order to prove the success of the model. Results have been derived both analytically and numerically
(for verification reasons) and are presented together with descriptions of the range of the involved parameters
that ensure: i) recent passage from deceleration to acceleration, ii) small outflow that do not violate black hole
evolution/mass, iii) nucleosynthesis bounds, and iv) universe age. Finally, section VI is dedicated to the conclusions.



\section{The framework: brane cosmology with 5-dim bulk energy exchange}

We begin with a model described by a 5-dim Einstein-Hilbert action with
matter and a 5-dim cosmological constant $\Lambda $ plus the contribution
describing the brane
\begin{equation}
S\!=\!\!\int \!d^{5}x\sqrt{-g}\,(M^{3}R-\Lambda +\mathcal{L}_{B}^{mat})+\int \!d^{4}x%
\sqrt{-h}\,(-V+\mathcal{L}_{b}^{mat}),
\end{equation}%
where $R$ is the Ricci scalar of the five-dimensional metric
$g_{AB}$ ($A,B=0,1,2,3,5$) and $h$ is
the induced metric on the 3-brane. We identify $\left( x,z\right) $ with $%
\left( x,-z\right) $, where $z\equiv x_{5}$, in order to impose the usual
$%
\mathbb{Z}
_{2}$ reflection symmetry of the AdS slice. Following the
conventions of \cite{RS}, we extend the bulk integration over the
entire
interval $\left( -\infty ,\infty \right) $. $\mathcal{L}_{B}^{mat}\ $and $%
\mathcal{L}_{b}^{mat}$ are the bulk and brane matter contents respectively. $%
M$ is the five-dimensional Planck mass. The quantity $V$ can include
the brane tension as well as quantum contributions to the
four-dimensional cosmological constant.

In order to search for cosmological solutions we consider the corresponding
form for the metric
\begin{equation}
ds^{2}=-n^{2}\left( t,z\right) dt^{2}+a^{2}\left( t,z\right) \gamma
_{ij}dx^{i}dx^{j}+b^{2}\left( t,z\right) dz^{2}\,,  \label{bulk
metric}
\end{equation}%
where $\gamma _{ij}$ is a maximally symmetric 3-dimensional metric with $%
i,j=0,1,2,3$ (we use $k=-1,0,1$ to parameterize the spatial curvature).
The five-dimensional Einstein equations are $G_{MN}=\frac{1}{%
2M^{3}}T_{MN}$, where $T_{MN}$ is the total energy momentum tensor, i.e.
\begin{eqnarray}
&&\!\!\!\!\!\!\!\!\!\!\!\!\!T_{N}^{M}=T_{N}^{M}\!\mid_{v,b}+T_{N}^{M}\!\mid_{m,b}+T_{N}^{M}
\!\mid_{v,B}+T_{N}^{M}\!\mid_{m,B}
\\
&&\!\!\!\!\!\!\!\!\!\!\!\!\!T_{N}^{M}\!\mid _{vac,b}=\frac{\delta
\left( z\right) }{b}diag\left( -V,-V,-V,-V,0\right)\nn
\\
&&\!\!\!\!\!\!\!\!\!\!\!\!\!T_{N}^{M}\!\mid _{vac,B}=diag\left(
-\Lambda ,-\Lambda ,-\Lambda ,-\Lambda ,-\Lambda \right)
\\
&&\!\!\!\!\!\!\!\!\!\!\!\!\!T_{N}^{M}\!\mid _{matter,b}=\frac{\delta
\left( z\right) }{b}diag\left( -\rho ,p,p,p,0\right)\nn
\\
&&\!\!\!\!\!\!\!\!\!\!\!\!\!T_{N}^{M}\!\mid _{matter,B}=diag\left(
0,0,0,0,T_{5}^{5}\right)\!+\!\left(\!\!
                           \begin{array}{cc}
                             \mathbb{O} & T^{0}_{5} \\
                             \frac{-n^2}{b^2}T^{0}_{5} & \mathbb{O} \\
                           \end{array}
                         \!\right)
\!\!.
\end{eqnarray}
Here, $T_{N}^{M}\!\mid _{m,b}$ denotes the energy-momentum tensor of
the brane perfect cosmic fluid and $\rho $, $p$ are its energy
density and pressure respectively. In our approach
$T_{MN}\!\mid_{m,B}$ gets non zero contributions from the presence
of flows from the brane. The off-diagonal contribution $T^{0}_{5}$
expresses the brane-bulk energy exchange flow of gravitons, while
the $T_{5}^{5}$ component expresses the corresponding pressure along
the fifth dimension. In order to keep predictability we seek to
derive a solution that is largely independent of the bulk dynamics.
Thus, any other existing bulk field contribution is considered
negligible. In addition, a small energy exchange from the brane
(true in the proposed mechanism) keeps the bulk largely unperturbed.
The set of the Einstein equations at the location of the brane is
\begin{eqnarray}
&&\!\!\!\!\!\!\!\!\!\overset{\mathbf{.}}{\rho}+3\frac{\overset{.}{a}_{o}}{a_{o}}\left(
\rho+p\right) =-\frac{2n_{o}^{2}}{b_{o}} \ T_{5}^{0}
\label{systema1}
\\
&&\!\!\!\!\!\!\!\!\!\frac{1}{n_{o}^{2}}\Big(
\frac{\ddot{a}_{o}}{a_{o}}\!+\!\Big( \frac{\overset{.}
{a}_{o}}{a_{o}}\Big)
^{2}\!-\!\frac{\overset{.}{a}_{o}}{a_{o}}\frac{\overset{.}
{n}_{o}}{n_{o}}\Big) +\frac{k}{a_{o}^{2}} =\frac{1}{6M^{3}}\Big(
\Lambda +\frac{V^{2}}{12M^{3}}\Big)\nn \\  &&
-\frac{1}{144M^{6}}\Big(V(3p-\rho) +\rho (3p+\rho)\Big)
-\frac{1}{6M^{3}}T_{5}^{5}\ . \label{systema2}
\end{eqnarray}
\newline
Dots indicate derivatives with respect to $t$. We indicate by the
subscript \textquotedblleft $o$" the value of various quantities on
the
brane and $T_{05\text{, }}T_{55\text{ }}$ are the 05 and 55 components of $%
T_{MN}$ evaluated on the brane.

Since we are interested in a model that reduces to the Randall-Sundrum
vacuum \cite{RS} in the absence of matter we require the bulk cosmological
constant and the brane tension to satisfy $\Lambda+\frac{1}{12M^{3}}V^{2}=0$.

It is convenient to employ a coordinate frame in which
$b_{o}=n_{o}=1$ in the above equations. This can be achieved by
using Gauss normal coordinates with $b\left( t,z\right) =1$ and by
going to the temporal gauge on the brane with $n_{o}=1$. Thus, using $\beta \equiv M^{-6}/ 144$ and $\gamma
\equiv V \beta $ and omitting the subscript $o$ for convenience in the
following, we take
\begin{eqnarray}
&&\overset{\cdot }{\rho }+3\left( 1+w\right) H \rho =-T \label{simple1} \\
&&q = 1 + H^{-2} \frac {k}{a^{2}} +H^{-2} (3w-1)\gamma \rho+  \nonumber \\
&&\quad\quad\quad+H^{-2} (3w+1)\beta \rho ^{2}\ +H^{-2} \sqrt{\beta}\ \Pi . \label{simple2}
\end{eqnarray}
Here, $q$ is the usual deceleration
parameter $q=-\frac{\ddot{a}}{a}H^{-2}$, $p=w\rho $ and $T=2T_{5}^{0},$ $\Pi =2T_{5}^{5}$ are the discontinuities
of the zero-five and five-five components of the bulk energy-momentum tensor respectively.
It is obvious from Eqs. (\ref{simple1}), (\ref{simple2}) that the only way to pass from a deceleration era to an
accelerated cosmic phase in a flat universe ($k=0$) is the case that the dark pressure term $\Pi$ becomes
negative at some moment in the cosmic history. This is actually what happens in the studied proposal where
dark pressure from zero acquires a non-zero negative value due to the leakage towards the extra dimension.
As we will explain in section III this leakage causes a negative pressure in every galactic center resulting
to a total value for $\Pi$ capable to produce the observed cosmic acceleration.

Defining now an auxiliary quantity $\psi $ by
\begin{equation}
\frac{\ddot{a}}{a}=-\left( 2+3w\right) \beta \rho ^{2}-\left( 1+3w\right)
\gamma \rho -\sqrt{\beta }\ \Pi -\psi +\lambda \quad , \label{accel}
\end{equation}
we can rewrite equations (\ref{systema1}), (\ref{systema2}) in the
equivalent form
\begin{eqnarray}
&&\overset{\cdot }{\rho }+3\left( 1+w\right) \frac{\overset{\cdot
}{a}}{a}\,\rho =-T  \label{systemb1}
\\
&&\frac{\overset{\cdot}{a}^{2}}{a^{2}}=\beta\rho^{2}+2\gamma\rho-\frac{k}{a^{2}%
}+\psi+\lambda  \label{systemb2}
\\
&&\overset{\cdot }{\psi }+4\frac{\overset{\cdot }{a}}{a}\psi =2\beta
\Big(
\rho +\frac{\gamma }{\beta }\Big) T-2\sqrt{\beta }\,\frac{\overset{\cdot }{a}%
}{a}\ \Pi \,. \label{systemb3}
\end{eqnarray}

Here, the effective cosmological constant on the brane $\lambda
=\left( \Lambda +V^{2}/ 12M^{3}\right) / 12M^{3}$, as we have
mentioned before, it will be set to zero, but for the time being we
leave it intact. Additionally, $\gamma=\frac{4\pi G_{N}}{3}$ in
order to recover standard 4-dimensional gravity.

In the special case of no-exchange $\left(
\Pi =0,\,T=0\right) ,\;\psi $ represents the dark radiation $\frac{\mathcal{C}}{a^4}$,
reflecting the non-zero Weyl tensor of the bulk.

In order to study further cosmic acceleration, it is convenient to
consider the set of differential equations (\ref{systemc1}),
(\ref{systemc2}) for $q, \rho$, equivalent to the last system of
equations (\ref{systemb1}), (\ref{systemb2}), (\ref {systemb3})
\begin{eqnarray}
&&\!\!\!\!\!\!\frac{dq}{da}=\frac{2}{a}q(q+1)+H^{-2}\Big[ 2\left(
2+3w\right) \beta \rho \frac{d\rho }{da}+(1+3w)\gamma \frac{d\rho
}{da}\nn
\\
&&\quad\quad\quad\quad\quad\quad\quad\quad\quad\quad\quad\quad\quad+\frac{d\psi
}{da}+\sqrt{ \beta }\frac{d\Pi }{da}\Big] \label{systemc1}
\\
&&\quad\quad\quad\quad\frac{d\rho }{da}=-\frac{1}{a}\left[
3(1+w)\rho +T\ H^{-1}\right]\,, \label{systemc2}
\end{eqnarray}
where we should replace everywhere $\psi $ and $d\psi /da$\ by \be
\psi =-(2+3w)\beta \rho ^{2}-(1+3w)\gamma \rho +H^{2}q-\sqrt{\beta
}\ \Pi +\lambda  \label{psi} \ee \be \frac{d\psi
}{da}=\frac{1}{a}\left[ -4\psi +2\beta \Big( \rho +\frac{\gamma
}{\beta }\Big) \ T\ H^{-1}-2\sqrt{\beta }\ \Pi \right] \ee and we
substitute $H^{2}$ with the help of
\begin{equation}
H^{2}=\frac{(3w+1)\beta \rho ^{2}+(3w-1)\gamma \rho +\frac{k}{a^{2}}%
-2\lambda +\sqrt{\beta }\ \Pi }{q-1}\,.  \label{constraint}
\end{equation}
It is worth mentioning
that both $\Pi $ and $T$, according to the proposed mechanism,
depend on the astrophysical properties of black holes.

We are going to distinguish two cases in our numerical study of the system
(\ref{systemc1}), (\ref{systemc2}). Both are consistent with the
details of the involved phenomena. In the first case, which is valid for a study of the recent
cosmological time period, i.e. $z$ close to 0, we assume a constant
number of typical galactic black holes and therefore, the dark
pressure $\Pi $ can be modeled to be analogous to a known constant
$\varpi $ times the inverse Hubble volume, $\Pi =\varpi H^{3}$ (see
section III for justification). The same holds also for the
outflow that can be approximated to be analogous to a known
constant $\tau $ times the inverse Hubble volume, $T=\tau H^{3}$.
Therefore,
the derivative $\frac{d\Pi }{da}$ that appears in (\ref
{systemc1}) should be replaced as follows
\begin{equation}
\frac{d\Pi }{da}=-\frac{3\varpi H^{3}(q+1)}{a}.
\end{equation}
Finally, $H$ can be found solving exactly the cubic equation
\begin{equation}
-\sqrt{\beta }\,\varpi H^{3}+(q-1)H^{2}\!=\!(3w+1)\beta \rho
^{2}+(3w-1)\gamma \rho +\frac{k}{a^{2}}-2\lambda.
\end{equation}
The above cubic equation for all realistic parameters has always two real positive roots associated with
expanding universe and one negative leading to contracting
cosmological solutions. These two roots for $H$ are given by
\begin{eqnarray}
&&H=s_{1}+s_{2}-A/3\qquad \text{or} \label{H} \\
&&H=-\frac{1}{2}(s_{1}+s_{2})-\frac{A}{3}-i\frac{\sqrt{3}}{2}(s_{1}-s_{2})
\,,
\end{eqnarray}
where
\begin{eqnarray}
&&\!\!\!\!\!\!\!s_{1}=\Big[ \eta +(\vartheta ^{3}+\eta
^{2})^{1/2}\Big] ^{1/3}\,\,,\,\, s_{2}=\Big[ \eta -(\vartheta
^{3}+\eta ^{2})^{1/2}\Big]^{1/3}\nn \\ && \!\!\!\!\!\!\!A=
\frac{1-q}{\sqrt{\beta }\,\varpi}\,\,\,,\,\,\, \vartheta
=-\frac{1}{9}A^{2}\nn\\&& \!\!\!\!\!\!\!\eta
\!=\!-\frac{1}{2}\frac{(3w\!+\!1)\beta \rho
^{2}\!+\!(3w\!-\!1)\gamma \rho +\frac{k}{a^{2}}\!-\!2\lambda
}{\sqrt{\beta }\varpi }\!-\! \frac{1}{27}A^{3}.
\end{eqnarray}
Now, the system of differential equations (\ref{systemc1}), (\ref{systemc2})
can easily be solved numerically. Thus, it is possible to test if the
measured cosmic acceleration can be produced from some sensible values of $T,\Pi$ according to our scenario.

In the second case which is more general, we do not assume a
constant number of black holes since we are interested to include
the dependence of the total mass density of black holes on the scale
factor evolution. In this way it will be possible to describe the
cosmological behaviour for redshifts far away
from $z=0$. The proposed mechanism suggests, in this second case, that $\Pi =\widehat{
\varpi }\ \rho _{BH}$ and $T=\widehat{\tau }\ \rho _{BH}$, where\ $\widehat{
\varpi }$, $\widehat{\tau }$ are known constants and\ $\rho _{BH}$
is the density of the relevant black holes which is a function of
the scale factor. Now, the derivative $\frac{d\Pi }{da}$ that
appears in (\ref{systemc1}), equals
\begin{equation}
\frac{d\Pi }{da}=\widehat{\varpi }\ \frac{d\rho _{BH}}{da}\,.
\end{equation}
Since it is possible to know estimations concerning the evolution of
$\rho _{BH}$ as a function of $a$ (see section V), it
is possible to solve numerically the system (\ref{systemc1}), (\ref{systemc2}).
In this second case, there is no need to solve any cubic equation
since we can estimate $H$ from (\ref{constraint}).

Let's now see if we can learn from these dynamic equations something
about the value of dark pressure we need to have in
order to explain cosmic acceleration. Equation (\ref{constraint})
provides a constraint that the cosmic acceleration should
satisfy. From it, we can determine the current value of $\Pi $ as a
function of the present values \bea
&&\!\!\!\!\!\!\!\!\!\!\!\!\!\!\!\!\!\!\!\!\!\!\!\!\!\!\!\!\!\Pi _{0}
=\Big(\!\!-1+q_{0}+\frac{\Omega _{m,0}}{2}\Big)\ \beta
^{-1/2}H_{0}^{2} \nn
\\
&&\Leftrightarrow \,\,\,\,\,\, q_{0} =1-\frac{\Omega _{m,0}}{2}+\Pi
_{0}H_{0}^{-2}\ \beta ^{1/2} \label{q0}, \eea where we have replaced
$\rho =\Omega _{m}\rho _{cr}$ and $k=0,$ $\lambda =0,w=0$. Note
that the current value of the deceleration parameter $q_{0}$ depends only on $\Pi _{0},H_{0}^{2},\Omega _{m,0}.\ $
Now, if we set $\rho _{0}\simeq \frac{1}{3}\rho_{cr, 0} $ and for example $q_{0}=-1$,  we get
\begin{equation}
\Pi _{0}\simeq-\frac{11}{6}\ H_{0}^{2}\ \beta ^{-1/2}\ .  \label{current po}
\end{equation}
Such negative values can be easily realized in our scenario, arising
from the proposed outflow mechanism. Therefore, it is possible
without solving the differential equations to check the efficiency
of the proposed mechanism using the value of $\Pi$ which is fully
determined from the involved astrophysical parameters and the value
of $M$. In this way, someone can be convinced that a small outflow
in each of the galactic black holes suffices to result to the
required amount of the total dark pressure and consequently to the
measured cosmic acceleration. Intuitively, one could say that the
geometry of the membrane universe is such that the negative dark
pressure ``stretches'' this membrane and causes acceleration.

Since $q$ is not directly measured, we have to express it as a
function of the ratios of cosmological matter density to critical
density and dark energy density to critical density. We define\qquad
\begin{equation}
\Omega _{m}=\frac{2\gamma \rho }{H^{2}}=\frac{\rho }{\rho
_{cr}},\quad \Omega _{\lambda }=\frac{\lambda }{H^{2}},\quad \Omega
_{k}=-\frac{k}{ a^{2}H^{2}}
\end{equation}
and for the dark energy part
\begin{equation}
\Omega _{DE}=\frac{\beta \rho ^{2}+\psi }{H^{2}}=\frac{\rho _{DE}}{\rho _{cr}}\,.
\end{equation}
Therefore, Eq. (\ref{systemb2}) gives
\begin{equation}
\Omega _{m}+\Omega _{DE}+\Omega _{\lambda }+\Omega _{k}=1\,.
\end{equation}
Finally, the deceleration parameter can be found from
\begin{equation}
q=\Omega _{DE}+\sqrt{\beta }\,\Pi\, H^{-2}+(1+3w)\,\frac{\Omega _{m}}{2}\,\Big(\!1+\frac{%
\beta H^{2}}{2\gamma ^{2}}\Omega _{m}\Big)-\Omega _{\lambda }\,.
\label{pum}\end{equation}
This last equation is going to provide us the initial condition for $%
q_{0}=q(z=0)$ given that $\Pi_{0}$ is provided by the astrophysical parameters, either in the first or second
case discussed above. Now, the system of differential equations (\ref{systemc1}), (\ref{systemc2} can be solved
using the initial conditions $q_{0}$, $\rho_{0}=\rho (z=0)$.

Another useful quantity used in physical cosmology is the
coefficient $w_{DE}$ of the equation of state of the dark energy. In our case the dark energy density encodes the
density required to represent the energy exchange. According to \cite{linder}, $w_{DE}$ is given by
\begin{equation}
w_{DE}=-1-\frac{1}{3} \frac{d}{d\ln a} \ln \Big{(} \frac{H^2}{H^2_{0}}- \frac {\Omega_{m,0}}{a^3} \Big{)}.
\label{wDE}
\end{equation}
It is straightforward to prove that
\begin{eqnarray}
&&\!\!\!\!\!\!w_{DE}\!=\!\frac{1}{\frac{H^{2}}{H_{0}^{2}}\!-\!\frac{\Omega_{m,0}}{a^{3}}}{\Big{\{}}
\!\Big{(}\frac{\Omega_{DE}}{3}\!+\!w\Omega_{m}\!+\!\frac{1\!+\!3w}{6}\frac{\beta}{\gamma
^{2}}\Omega _{m}^{2}H^{2}\Big{)}\frac{H^{2}}{H_{0}^{2}}\nn \\
&&\,\,\,\,\,\,\,\,\,\,\,\,\,\,\,\,\,\,\,\,\,\,\,\,\,\,\,\,\,\,
\,\,\,\,\,\,\,\,\,\,\,\,\,\,\,\,\,\,\,\,\,\,\,\,
\,\,\,\,\,\,\,\,\,\,\,\,\,\,\,\,\,\,\,\,\,\,\,\,\,\,+\frac{2\sqrt{\beta
}\,\Pi }{3H_{0}^{2}}{\Big{\}}}\,, \label{wDE}
\end{eqnarray}
and therefore, the today value (for $w=0$) is
\begin{equation}
w_{DE,0}=\frac{1}{3}+\frac{\beta}{6\gamma^{2}}\frac{\Omega _{m,0}^{2}H_{0}^{2}}{\Omega _{DE,0}}+\frac{2\sqrt{\beta
}\,\Pi_{0}}{3\Omega _{DE,0}H_{0}^{2}}\,.
\end{equation}
The numerical value of $w_{DE,0}$ will be a prediction for our model.
This equation manifestly shows that a negative dark pressure term
can easily cause not only cosmic acceleration but also the crossing
of the $w_{DE}=-1$ phantom divide line. This was pointed out in a
different context in \cite{tamva}.

Finally, an alternative useful expression that can be derived from
equations (\ref{systemb1}), (\ref{systemb2}), (\ref{systemb3}) is
the following single differential equation for $k=0$ that depends on
the energy density and which can be easily solved numerically
\begin{eqnarray}
&&\!\!\!\!\!\!\!\!\!\!\!\!\!\!\!\!\!\!\!\frac{dq}{d\rho }=2q\left(
q+1\right) Z^{-1}-3H\sqrt{\beta}\,(q+1)\varpi \
Z^{-1}\nn\\
&&\!\!\!\!\!\!\!+H^{-2}\Big[ 2\left( 2+3w\right) \beta \rho
+(1+3w)\gamma
-4\psi \ Z^{-1}\nn \\
&&\,\,\,\,\,\,\,\,\,\,\,+2\beta \Big(\rho +\frac{\gamma }{\beta
}\Big) T Z^{-1}X-2\sqrt{\beta }\ \Pi \ Z^{-1}\Big],
\end{eqnarray}
where
\begin{eqnarray}
Z &=&-3(w+1)\rho -T\ X \\
X &=&(\beta \rho ^{2}+2\gamma \rho -\frac{k}{a^{2}}+\psi +\lambda )^{-1/2}
\end{eqnarray}%
and we replace everywhere $H$ and $\psi$ from equations (\ref{psi}), (\ref{H}) or (\ref{constraint}).

As mentioned, in the Randall-Sundrum model the effective
cosmological constant $\lambda $ vanishes, and this is the value we
assume in the rest of the paper. We also set $k=0$ since we are
interested on flat universes. Finally, since we are analyzing the
cosmic acceleration after the large scale structure of the universe
we set $w=0$.

\section{A novel phenomenon: brane-bulk energy exchange inside galactic core black holes and/or
galactic halo black holes}

It is quite natural in the
framework of brane cosmologies to expect a small energy exchange of
our brane universe with the bulk space. This energy exchange
phenomenon is a high energy phenomenon. The channels for energy
exchange \textquotedblleft open" when the relevant energies reach
the relatively low five-dimensional fundamental Planck energy scale $M$. In the
cosmological context regions where such high energy phenomena could
occur are not as many.

In general, we would expect a brane-bulk energy exchange through:

\begin{enumerate}
\item High energy interactions in some accretion disks and more importantly
inside galactic centres/galactic black holes leading to energy loss to the
bulk due to the production of gravitons from high energetic accelerated
particles.

\item Gravitational attraction of a portion of the gravitons that were
escaped into the bulk or gravitational accretion of bulk matter to brane
black hole.

\item Attraction of bulk matter from the whole brane.

\item Decay of very massive scalars and/or fermions.
\end{enumerate}

The third type of exchange can be seriously studied only if the bulk matter
content is known in detail, see for example \cite{ovrut0}, and consequently
only if we are sure about the geometry of the bulk space, its anisotropies
and the motion of our brane in it. Various different approaches to describe
bulk dynamics/matter can be found in \cite{kiritsis}, \cite{kraus}, \cite%
{tetradis}, \cite{tetradis2}, \cite{ovrut}, \cite{kirske}.

The fourth exchange mechanism \cite{dubovsky} works only for very
massive particles like light supersymmetric particles with masses
above 1TeV. This option to produce the measured acceleration is
not very generic (see \cite{ichicki2}).

The second exchange mechanism regarding attraction of the escaping gravitons contributes to dark pressure,
but only to a small amount at late times \cite{langlois}, for which
we are interested in. On the other hand, the attraction of bulk matter from the
gravitational field of brane black holes may be not negligible for
significant values of bulk matter density. Nevertheless, since we are
interested to investigate energy exchange without losing predictability we
have assumed that the matter energy density of bulk fluid is small or zero.


Our proposed mechanism considers a
homogeneous distribution of galactic black holes on a brane. The
study of a Swiss cheese-like brane world model with bulk energy exchange
in each black hole (extension of work \cite{gergely}) certainly would be a more precise modeling.
The case is analogous to a swiss-cheese like brane world model
with Schwarzschild-de Sitter black holes finally resulting to a dust FRW cosmology with an overall
cosmological constant arising from each black hole contribution.
However, the modifications of this more precise modeling are
expected to be small. Therefore, the total brane-bulk energy exchange under consideration will be
\begin{equation}
T=T_{e}\ ,
\end{equation}
where $T_{e}$ represents the outflow of energy due to the production of
escaping gravitons from high energetic particles inside BHs. Note also that in the present paper we will not
consider the possibility of an existing considerable amount of primordial black holes today, and
therefore we will not study further scenarios with such black holes.

Last years it became evident that every nearby massive galaxy
possesses a central black hole with mass proportional to that of the
galaxy spheroid. This implies that they also possess an Active
Galactic Nuclei (AGN) \cite{alexander}. In addition, there are
evidences for the existence of a large amount of extra-galactic
exposure at TeV energies \cite{herr}, \cite{zdrziarski} and some of
it can be associated to the presence of galactic black holes and
galactic core supermassive black holes.

It is certainly a safe assumption that in the accretion discs and more
importantly in the interiors of galactic black holes and galactic core
supermassive black holes various particles as electrons and protons can be
thermalised/accelerated to energies around $M$ or above. Particle
acceleration starts in the accretion discs outside the horizon and increases
as the particle crosses it. Consequently, particle collisions become capable
to produce gravitons escaping to the bulk space.

Moreover, assuming a black hole with physics that respects unitarity in its interior, it is acceptable to
use the picture of an effective quantum fluid that fills the black hole and does not concentrate at the singular
center (otherwise there will be information loss from the exactly thermal Hawking radiation). This effective fluid
may be on a high temperature below or close to $M.$ At these energies it is
possible \cite{langlois}, \cite{Hebecker} to obtain rapid energy
favored production of bulk gravitons from collisions of energetic
brane matter. In a hot plasma the production rate per 3-volume is
the thermal average of the cross section times the lost energy of
the particles. Therefore, the total energy loss rate due to bulk
graviton radiation is \cite{Hebecker}, \cite{langlois}%
\begin{equation}
\Delta \overset{\cdot }{\rho }_{pls}=0.112\frac{\Theta
^{4}}{2M^{3}}\rho _{pls}=0.112\ g_{\ast }\frac{\pi
^{2}}{60}\frac{\Theta ^{8}}{M^{3}}\, , \label{escapeplasma}
\end{equation}%
where $\Theta $ is the temperature and $\rho _{pls}$ is the total energy density of the hot regions.
The second equation in (\ref{escapeplasma}) is derived assuming a relativistic
plasma with $g_{\ast }=106.75$ the effective number of the relativistic degrees of
freedom.

In order to proceed to a rough estimation of the mean outflow energy rate,
an effective mean black hole plasma energy-mass density $\rho _{pls}^{BH}~$%
is assumed inside brane black holes expressed with the help of an
effective mean temperature $\Theta _{mean}.$ The following expression will be
used
\begin{equation}
\rho _{pls}^{BH}\simeq g_{\ast }\frac{\pi ^{2}}{30}\Theta _{mean}^{4}\,.
\end{equation}
Now, let $\Delta\dot{\rho}_{tot}$ be the leakage of energy from the total volume of the warm plasma of a black hole. In order to evaluate $T_{e}$ we have to add all these leakages from all galactic halo black holes and all black holes at the galactic central regions and divide with the Hubble volume $H^{-3}$, thus $T_{e}=H^{3}\sum\Delta\dot{\rho}_{tot}$. Since the total volume in the universe of warm black hole plasma is $N_{BH}V_{BH}$, we get
\begin{eqnarray}
&&\!\!\!\!\!\!\!\!\!\!\!\!\!\!\!T_{e} \simeq 0.112\ g_{\ast
}\frac{\pi ^{2}}{60}\frac{\Theta _{mean}^{8}}{
M^{3}}[N_{haloBH}V_{haloBH}\nn
\\&&\quad\quad\quad\quad\quad+N_{coreBH}V_{coreBH}]\,H^{3}\quad \text{or}  \notag \\
&&\!\!\!\!\!\!\!\!\!\!\!\!\!\!\!T_{e} \simeq
\frac{0.112}{2M^{3}}\Theta _{mean}^{4}[N_{haloBH}M_{haloBH}\nn
\\&&\quad\quad+N_{coreBH}M_{coreBH}]\,H^{3}=\tau
H^{3}\quad \text{or}  \label{T0} \\
&&\!\!\!\!\!\!\!\!\!\!\!\!\!\!\!T_{e} \simeq
\frac{0.112}{2M^{3}}\Theta _{mean}^{4}(\rho _{haloBH}+\rho
_{coreBH})\nn
\\&&\!\!\!\!\!\!\!=\ \widehat{\tau }\,\,(\rho
_{haloBH}+\rho _{coreBH})=\widehat{\tau } \,\, \rho _{BH} \,.
\label{T_2ndcase}
\end{eqnarray}%
We assume the existence of $\ N_{haloBH}$ galactic halo black holes
with mean mass $ M_{haloBH}$ and $N_{coreBH}$ galactic central
regions carrying a supermassive black hole with a mean value equal
to $M_{coreBH}.$ Since the mean value of mass density $\rho
_{pls}^{BH}$ are very different among a typical halo black hole and
a typical supermassive core black hole, we substitute $
V_{haloBH}=M_{haloBH}(\rho _{pls}^{hBH})^{-1}$and
$V_{coreBH}=M_{BHcore}( \rho _{pls}^{cBH})^{-1}$. Although for
simplicity in the above formulae the temperature appears as a common
mean value, in reality $\Theta _{mean}$ can be different between
halo and core black holes, something that has been considered in the
numerical study of the solutions.

The magnitude of the three dimensional pressure inside the black
hole is equal to the magnitude of the pressure of the effective
fluid. Since our collapsing fluid is not an ideal fermi gas, we
adapt an index $\widehat{\gamma }$ for determining the three
dimensional pressure in the interior of both halo and core black
holes, i.e.
\begin{equation}
p_{pls}^{BH}=\xi (\rho _{pls}^{BH})^{\widehat{\gamma }}\,.
\end{equation}
The constant $\xi ~$\ is determined by the thermal characteristics
of the fluid and it can also be understood as a measure of the ratio
of pressure to energy density at the center of black hole.
Since the aim is to determine the dark pressure towards the fifth dimension we should divide the three dimensional
pressure with the characteristic kinetic length scale  $L$ of the plasma towards the bulk $p_{pls}^{BH}/L$.
This length $L$ has been proven in \cite{langlois} that is $L=\frac{M^3}{\rho_{pls}}$. This is the reason why in
Eq. (\ref{escapeplasma}) the outflow is analogous to $\frac{\rho_{pls}^2}{M^3}$.
Therefore we get
\begin{equation}
\Pi^{BH}=-\xi \frac{(\rho_{pls}^{BH})^{\widehat{\gamma }+1}}{M^3}.
\end{equation}
Note that the last expression for $\widehat{\gamma }=1$ reduces to the dark pressure estimated in \cite{langlois}.

The phenomenon under discussion most importantly
results to the appearance of a negative pressure orthogonal to the
fifth dimension. At the position of the brane the five-dimensional pressure $\Pi =2T^{55}$ equals the momentum
flux carried from the bulk to the brane. Because of momentum conservation this pressure equals the opposite of
the momentum flux carried by the escaping gravitons from the brane to the bulk. Therefore, $\Pi<0$. This negative
sign is a subtle point missed in the analysis in \cite{langlois}.

Finally,
\bea &&\!\!\!\!\!\!\!\!\Pi=-\xi \lbrack \frac{(\rho
_{pls}^{hBH})^{\widehat{\gamma }+1 }}{M^3}N_{haloBH}V_{haloBH}\nn
\\
&&\quad\quad\quad\,\,\,\,\,+\frac{(\rho
_{pls}^{cBH})^{\widehat{\gamma }+1 }}{M^3} N_{coreBH}V_{coreBH}]\ H^{3}  \notag
\\
&&\!\!=-\xi \lbrack \frac{(\rho _{pls}^{hBH})^{\widehat{\gamma }}}{M^3}N_{haloBH}M_{haloBH}\nn
\\
&&\,\,\,\,\,\,\,\,\,\,\,\,\,\,\,\,\,\,\,\,\,\,\,+\frac{(\rho
_{pls}^{cBH})^{\widehat{\gamma } }}{M^3}N_{coreBH}M_{BHcore}]\ H^{3}\nn
\\
&&\!\!=\varpi H^{3}  \label{P0}
\\
&&\!\!=-\xi \lbrack \frac{(\rho _{pls}^{hBH})^{\widehat{\gamma }}}{M^3}\rho
_{haloBH}+\frac{(\rho _{pls}^{cBH})^{\widehat{\gamma }}}{M^3}\rho
_{coreBH}]\nn
\\
&&\!\!= \widehat{\varpi }\ \rho _{BH}\,.  \label{P_2ndcase} \eea
Equation (\ref{P_2ndcase}) holds for the case where the mass density
of the core black holes is dominant.

It should be noticed that the proposed outflow mechanism has no
similarity with scenarios that set the density of the plasma equal to the
density of the overall cosmological fluid which cools as the universe
expands. For example in \cite{Hebecker}, \cite{langlois} the whole universe has to be thermalised in
temperatures close to the
fundamental planck scale which is not true at late times of the
evolution. In our work the thermalized fluid is in the interiors of black holes and leaks
towards the bulk. Furthermore, the pressure $T_{55}$ of the fluid is not of
the same order with the $T_{05}$ leakage as in \cite{langlois} since a non ideal gas quantum fluid is
expected/assumed inside black holes.

Based on the above discussion we can directly see the connection of
the present energy density of the universe $\rho_{0}$ to the observed
dark energy. Namely, the outflow energy rate and the dark pressure
are \be T_{e,0}=0.112\ \frac{\Theta _{mean}^{4}
}{2M^{3}}\,\varepsilon\,\rho _{0}, \ee \be \Pi _{0}=-\xi
\Big(g_{\ast }\frac{\pi ^{2}}{30}\Theta
_{mean}^{4}\Big)^{\widehat{\gamma } }\frac{1}{M^3}\ \varepsilon \rho _{0}, \ee
while the current cosmic acceleration becomes \be
q_{0}=1-\frac{\Omega _{m,0}}{2}-\xi \Big(g_{\ast }\frac{\pi
^{2}}{30}\Theta _{mean}^{4}\Big)^{\widehat{\gamma
}}\frac {1}{12 M^6} H_{0}^{-2}\ \varepsilon \rho _{0}. \label{current
po2} \ee The quantity $\varepsilon $ is the portion of the present
black hole mass density $ \rho _{BH,0}$ relative to the present
cosmic mass density $\rho _{0}$. In section V it will be
demonstrated that even for the most conservative values of all the
involved parameters such negative values of $q_{0}$ can be achieved.

\vspace{0.4cm}

\section{Additional support of the proposed mechanism}

In this section various physical aspects of the proposed mechanism are presented.

\subsection{The proposed mechanism and the gravitational collapse on the
brane}

In this subsection estimations are presented concerning the
evolution of a spherical collapse in the brane scenario presented
above. Our goal is to describe quantitatively the expected behavior of temperature rise
as the collapse of a fluid proceeds. \textit{In our case, strong quantum gravity
corrections are not necessary} since our intention is to describe
the collapse up to the point where the outflow becomes significant.
This happens for temperatures close to the fundamental Planck scale
which can be relatively low.

The spherical gravitational collapse on a brane with a realistic
brane-bulk energy exchange will now be analyzed. The interior of the
collapsing spherical region undergoing an Oppenheimer-Snyder
collapse will be described by the brane cosmological metric
(\ref{bulk metric}) presented above, with nonzero $T_{05},T_{55}$.
Therefore the evolution has to be a contracting solution of the
system of the brane cosmological equations (\ref {systemb1}),
(\ref{systemb2}) and (\ref{systemb3}). Now, the energy density, the
dark radiation and the dark pressure concern the plasma in the
interior of black hole/collapsing region. Thus, the system of
differential equations that the evolution of the collapsing region
should respect is \bea
&&\!\!\!\!\!\!\!\!\!\!\!\!\overset{}{\overset{.}{\rho
}_{pls}}+3\left( \rho _{pls}+p_{pls}\right) \frac{\overset{\cdot
}{R}}{R}=-T_{pls}
\\
&&\!\!\!\!\!\!\!\!\!\!\!\!\frac{\overset{\cdot }{R}^{2}}{R^{2}}=\beta \rho _{pls}^{2}+2\gamma
\rho _{pls}-\frac{\kappa}{R^{2}}+\psi
\\
&&\!\!\!\!\!\!\!\!\!\!\!\!\overset{\cdot }{\psi }+4\frac{\overset{\cdot }{R}}{R}\psi =2\beta
\Big(
\rho _{pls}+\frac{\gamma }{\beta }\Big) T_{pls}-2\sqrt{\beta }\,\frac{%
\overset{\cdot }{R}}{R}\ \Pi _{pls}\,, \eea where $\kappa$ characterizes the spatial topology of the collapsing shell
(with most interesting case $\kappa=1$). Here, the scale factor
$R(t)$ of the collapse region is related to the proper radius $r$
from the center of the cloud through $r=R\chi/(1+\kappa\chi^{2}/4)$,
where $\chi$ is the comoving coordinate and the dot denotes a proper
time derivative. The energy outflow and dark pressure are given by
\bea &&\!\!\!\!\!\!\!\!\!\!\!\!\!\!\! T_{pls}\simeq \frac{1.68}{\pi
^{2}g_{\ast }}\frac{1}{M^{3}}\rho _{pls}^{2}\simeq 0.112\ g_{\ast
}\frac{ \pi ^{2}}{60}\frac{\Theta _{mean}^{8}}{M^{3}}
\\
&&\!\!\!\!\!\!\!\!\!\!\!\!\!\!\!\Pi _{pls}\!=\!-\xi\,\rho
_{pls}^{\widehat{\gamma }+1}\frac{1}{M^{3}} \!\simeq\!-\xi
\Big(g_{\ast }\frac{\pi ^{2}}{30}\Theta
_{mean}^{4}\Big)^{\widehat{\gamma }+1}\frac{1}{M^{3}}. \eea
Collapsing plasma has been assumed to have an equation of state deviated
from this of an ideal gas. The relation between the energy
density and the temperature (local thermodynamic equilibrium) is
given by the ansatz $\rho _{pls}\simeq g_{\ast }\frac{\pi
^{2}}{30}\Theta _{mean}^{4}\simeq \sigma\,
\Theta _{mean}^{4}$. Pressure is expressed as $p_{pls}=\xi \rho _{pls}^{
\widehat{\gamma }}$. Realistic quantum fluids can effectively be described by an
equation of state with deviations from ideal gas behavior
\cite{Bannur}.

Therefore, in order to study the temperature evolution as the
collapse continues, we have to find solution of the following system
of differential equations \be \overset{\cdot }{\Theta
}_{mean}+\frac{3}{4}\Theta _{mean}\Big( 1+\xi \sigma
^{\widehat{\gamma }-1} \Theta _{mean}^{4(\widehat{\gamma } -1)}\Big)
\frac{\overset{\cdot }{R}}{R}\ +0.014\frac{\Theta _{mean}^{5}}{
M^{3}}\!=\!0 \ee \be\frac{\overset{\cdot }{R}^{2}}{R^{2}}=\beta
\sigma ^{2} \Theta _{mean}^{8}+2\gamma  \sigma  \Theta
_{mean}^{4}-\frac{\kappa}{R^{2}}+\psi \ee \bea
&&\!\!\!\!\!\!\!\!\!\!\!\!\!\!\!\!\!\!\!\!\!\! \overset{\cdot }{\psi
}+4\frac{\overset{\cdot }{R}}{R}\psi =0.112 \beta\sigma \Big(\sigma
\Theta _{mean}^{4}+\frac{\gamma }{\beta }\Big) \frac{\Theta
_{mean}^{8}}{M^{3}}\nn
\\
&&\quad\quad\quad\quad\quad\quad+2\sqrt{\beta }\,\xi\, \sigma
^{\widehat{\gamma }+1 }\, \frac{\overset{\cdot }{R}}{R} \Theta
_{mean}^{4(\widehat{\gamma }+1)}\frac{1}{M^{3}}. \eea
The above system of the first and third
equation can be solved numerically without difficulties. This study
shows that for expected parameters $ \frac{\Theta _{mean}}{M}<1$ and
for $\overset{\cdot }{R}<0$, which is the case of spherical
collapse, we can get $\overset{\cdot }{\Theta }_{mean}>0$, which is
what we want to prove.
A typical solution of this system is shown in Fig. 1,
where time $t$ is measured in GeV$^{-1}$ and temperature $\Theta$ in
GeV.


\begin{figure}[h!]
\centering
\begin{tabular}{cc}
\includegraphics*[width=240pt, height=160pt]{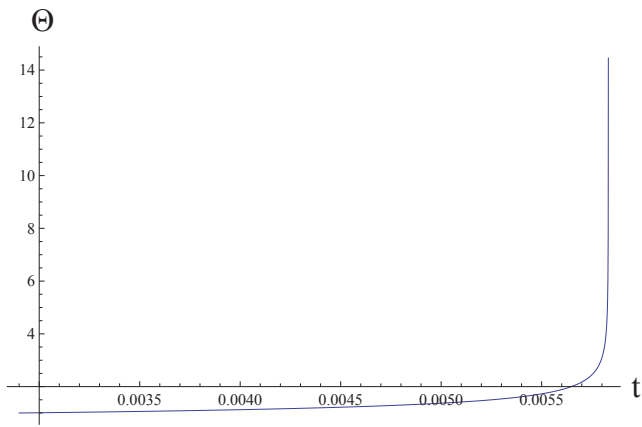}&%
\end{tabular}
\caption{Temperature rise during a collapse for $M=10^{4}$ GeV}
\end{figure}




Note that we are not interested here to find a static exterior for the
above described collapsing spherical region \cite{Pal}, \cite{harko}, \cite{harko2}.

\subsection{The proposed mechanism and the Hawking-like radiation}

An interesting work about the physics inside the forming horizon of
collapsing shells or the interior of black holes accreting matter is
this of Greenwood, Stojkovic \cite{greenwood}. In this work, Hawking
radiation was studied as seen by an infalling observer. Based on
functional Schrodinger formalism it is possible to calculate
radiation in Eddington-Filkenstein coordinates which are not
singular at the horizon. In these coordinates Hawking radiation does
not diverge on the horizon. The estimated occupation numbers at any
frequency, as measured by an observer crossing the horizon, were
found to increase as the distance from the black hole center
decreases. The spectrum is not thermal and therefore there is no
well-defined temperature measured by the observer. Although this
work does not refer to brane black holes, we expect similar
qualitative behavior for this case too. Therefore, the above
discussion suggests that an observer entering the horizon
encounters/interacts with more and more highly energetic particles
which can escape easily to the bulk or cause through their
interactions energy loss to the bulk. Estimations presented in
\cite{greenwood} are not valid for distances close to the black hole
centre where strong backreaction effects have to be considered.
However, in our case energy loss can start inside and near the horizon for temperatures close to $%
M$.

\subsection{The proposed mechanism and the Fuzzball approach}

A fluid description is certainly a phenomenological picture which is
traditionally followed in relativistic cosmology/astrophysics. Here
we will attempt to discuss microscopically the reason why such a
description can be based on fundamental physics. It is expected that
in reality in the interior and most certainly near the centers of
black holes the notion of classical spacetime is replaced by another
not well understood \textquotedblleft quantum" spacetime. The full
treatment is still unknown; nonetheless it is expected that an
\textit{effective} description of the black hole interior with the
help of a ``non perfect" fluid without infinite density could be a fair
approximation. Adopting this effective approach we assume that at
the center of the black hole, the density is
large but finite and equals to $\rho _{pls}(r=0).$ Therefore, if the
physics in the interior of astrophysical black holes was known, one
could in principle be able to reproduce an effective description
estimating
a mean value of the plasma density $\rho _{pls}^{BH}=\frac{4\pi }{V_{BH}%
}\int\limits_{0}^{R}\rho _{pls}(r)r^{2}dr$. The value of the effective
radial dependent plasma density $\rho _{pls}(r)$ as well as the effective
central finite value $\rho _{pls}(0)$ would then be determined by quantum
gravity.

Although in pure general relativity such an expression has no meaning since
the energy density becomes infinite and the spacetime description breaks at
the center, new ideas arising from string theory possibly allow an effective
quantum statistical description of the black hole interior. A promising
approach for addressing questions regarding physics inside black holes is
the fuzzball proposal \cite{mathur1}, \cite{mathur2}, \cite{mathur3}, \cite%
{mathur5}, \cite{mathur6}, \cite{skenderis}, \cite{bena}. According to this
view, the infinite \textquotedblleft throat" that a classical geometrical
description exhibits near the singularity is replaced by a long finite
throat which ends in a quantum fuzzy cap. The fuzzball conjecture claims
that the astrophysical black holes are described by microstates which all
behave like the ones that have been constructed for extremal black holes in
string theory. The bound states in string theory are not in general Planck
sized or string sized, but have a size that grows with the degeneracy of the
bound state. To make a big black hole a large number of elementary quanta
need to be placed together. Regarding the size of the bound state one may
think that this is equal to string or Planck scale $l_{pl}$. However, if
this was true we would get the traditional picture of brane black holes with
all matter placed at the singularity (then  hawking radiation becomes exactly
thermal leading to loss of unitarity). The correct picture is that the size
of the bound state increases with the number of quanta in the bound state.
In the fuzzball approach the size of the bound state $\Re \sim N^{a}l_{pl}$
has been proven to be equal to the black hole horizon radius that we would
find for the classical geometry which has the mass and charge carried by
these $N$ quanta. $N$ is some count of the quanta and $a$ depends on what
quanta are being bound together.

The fuzzball theory suggests two important
elements regarding the physics of the brane black holes interior.
First, the matter content is distributed all over the interior, a
fact that allows an effective description with a quantum statistical
fluid described by a non conventional equation of state. Second, some of the
quanta are free to tunnel into the bulk not due to Hawking black
hole evaporation but due to the absence of microstate horizons and
the brane-bulk geometry associated with a ``small" value of $M$.
Hawking radiation is due to fractional brane-antibrane
annihilations, while outflow is the result of tunneling of string
quanta of fractional and non fractional branes-antibranes towards
the bulk space.

Let us think in more detail what may happens inside a black hole. If
we increase the energy density of a collection of branes to very
large values, it becomes entropically favorable to produce a large
number of sets of mutually BPS branes and anti-branes. These branes
``fractionate" each other, resulting to entropy that grows more
rapidly as a function of energy compared to that of radiation or a
Hagedorn type string or brane gas. Therefore, in the case of
astrophysical black holes it is expected that after the beginning of
the collapse energy density grows and matter reaches a Hagedorn
phase of strings. Although this pressureless phase keeps its energy
nearly constant (there are already significant open outflow
channels) thanks to the continuing collapse the energy density
increases further. Finally, we end up to an even higher energy scale
phase with a soup of many fractional and less non fractional branes.

In the two charge system NS1-P bound state there is a string that
loops $n_{1}$ times around $S^{1}$ (radius $R$) with a momentum
charge $P$ which is bound to the string in the form of traveling
waves on the NS1 brane. The number of states that contribute more to
the entropy is approximately equal to $\exp (\sqrt{n_{1}n_{p}}).$
These states are fractional with a length $L_{T}$ equal to the
classic geometry horizon (if we add one more charge). These
fractional states have a low temperature/average energy (equal to
Hawking temperature if we add one more charge) given by
\begin{equation}
T_{H}=\frac{\sqrt{n_{1}n_{p}}}{L_{T}}\ ,
\end{equation}%
where the total length of the string is large and equal to%
\begin{equation}
L_{T}=2\pi R\ n_{1}
\end{equation}%
since in realistic astrophysical black holes $n_{1}$ can be very large.

However, in the black hole interior there are also fewer states with large
temperature/energy because 1) $L_{T}$ can be very small since $R$ is very
small, while $n_{1}$ is also small for non fractional states, 2) branes need
a large time (evaporation timescale) to fractionate to very large lengths.
These non fractional states tunnel immediately to the bulk space as long as
\begin{equation}
M\leq \frac{\sqrt{n_{p}}}{2\pi R\ \sqrt{n_{1}}}\quad .
\end{equation}%
Now the disappearing states to the bulk due to tunneling are continuously
replaced in the high energy density regions of the interior at the cost of
the collapsing matter's energy density.\ Thus, we have a non vanishing flow
of energy towards the bulk.

In the three charge system there are $n_{5}$ NS5 branes and $n_{1}$ NS1
branes that define a system with a momentum charge $P$. Therefore, the bound
system of these branes\ generate an \textquotedblleft effective
string\textquotedblright\ with a total winding number $n_{1}n_{5}$. All the
above discussion for the two charge system and all relevant expressions
remain the same replacing everywhere $n_{1}$ with $n_{1}n_{5}.$

Apart from the outflow originated by these states in the black hole
interior, there are two more outflow open channels. As we have previously
mentioned, a portion of collapsing matter is still in the string/brane gas
phase which is a very hot phase that certainly can leak to the bulk space.
In addition, there must be a non negligible outflow from the portion of the
collapsing matter that is between the string/gas phase and the electroweak
energy scale ($\sim $TeV) as long as its local temperature is close or
larger than $M$.

In summary, the reasoning that ensures outflow is the observation
that astrophysical black holes are not non-perturbative
configurations composed of wrapped strings or branes living at the
Planck regime or M-theory landscape. They are objects created
dynamically from collapsing matter initially respecting our U(1)
vacuum. This matter unavoidably gets compressed to smaller and
smaller volumes until it reaches very high energy scales where
outflow is not negligible and unavoidable.

To close this section, it is important to mention that although the
fuzzball proposal is helpful in order to understand the microscopic
processes of the outflow, the proposed mechanism operates based only
on two sensible requirements: first, the existence of
Schwarzschild-like black hole solutions on the brane with nonzero
$T^{0}_{5}$, proved in ref. \cite{fraction2} and used in a more general form
here, and second, the existence of a non conventional quantum fluid in the
interior of black holes or better the validity of an effective
description of the interior with such a fluid, something that sounds
natural since quantum states have to be important at the horizon,
otherwise thermal Hawking radiation would lead to information loss.

\section{Amount of produced cosmic acceleration}

The goal of the present work is to estimate for the proposed brane-bulk
energy exchange mechanism the amount of the produced present cosmic
acceleration for various values of the relevant parameters. This
section presents the numerical results of our study.

{\it{I. Numerical analysis with time-dependent black hole cosmic density}}. First we analyze the more general and interesting case of time dependent black hole cosmic mass density (referred in section II as second case) considering astrophysical estimates reported in \cite{A.Merloni}. With the help of them we can describe the core
black hole density evolution with the following relation (valid for
$z<2$)
\begin{equation}
\log _{10}(\rho _{coreBH})=-\mu \,z+\log _{10}(\rho _{coreBH}|_{z=0})\quad
. \label{rhoofz}\end{equation}%
Since $\rho _{coreBH}|_{z=0\text{ }}=4.3\!\times\!10^{5}M_{\odot
}Mpc^{-3}$
 is the current galactic core black hole matter density and $\rho
_{coreBH}|_{z=2}=1.5\!\times\!10^{5}M_{\odot }Mpc^{-3}$ is the density at redshift $%
z=2$ we obtain
\begin{equation}
\mu =\frac{\log _{10}(\rho _{coreBH}|_{z=0})-\log _{10}(\rho
_{coreBH}|_{z=2})}{2}\quad .
\end{equation}%
Equation (\ref{rhoofz}) shows that when the redshift $z$ decreases
(cosmic matter density decreases), the energy density of black holes
increases. Therefore, from (\ref{P_2ndcase}) we see that the
absolute value of dark pressure increases for the late stages of
cosmic evolution. Equations (\ref{pum}), (\ref{wDE}) show that
$q,w_{DE}$ get progressively negative values.  Based on the
expression (\ref{rhoofz}) it is possible to estimate the dependence
on the scale factor of dark radiation and dark pressure from Eqs.
(\ref{T_2ndcase}),
(\ref{P_2ndcase}). The numerical investigation of (\ref{systemc1}), (\ref%
{systemc2}) reveals that for a wide range of the parameters $\hat{\gamma}, M$
it is always possible to find a range for the mean temperature $\Theta _{mean}$
that results to cosmological solutions with current cosmic
acceleration $q<0$, with $w_{DE}$ around -1, and equally importantly
with a deceleration era that only currently becomes acceleration.
Table 1 presents some representative results, while Fig. 2 shows the
evolution of the deceleration parameter $q$ and $w_{DE}$ for
$\hat{\gamma}=0.05$, $M=50$TeV. Results in Table 1 reveal that it is possible to get cosmic acceleration for reasonable values of $M$ and $\Theta_{mean}$. For values $M>10^3$TeV there is a need for very large temperatures in the interior of black holes. Another remark is that the index $\hat{\gamma}$ has to be less than unity and this would be connected with the physical properties of the assumed quantum fluid in the interior of the black holes. Also note that the values of $\Pi$ are orders of magnitude larger than $T$.

It is also worth mentioning that both $T$, $\Pi$ are zero
before large scale structure since black holes have not appeared yet. Only after the large scale structure and the growth of
a significant population of astrophysical black holes the mechanism
is able to result to cosmic acceleration. The latter observation
provides a natural solution to the coincidence problem.

\begin{center}
\begin{table*}
\begin{tabular}{|c|c|c|c|c|c|}
\hline\hline \multicolumn{1}{||c|}{assumption} &
\multicolumn{1}{||c|}{assumption} &
\multicolumn{1}{||c|}{assumption} & \multicolumn{1}{||c|}{output} &
\multicolumn{1}{||c|}{output} & output \\ \hline\hline
$\hat{\gamma}$ & $M$(TeV) & \,\,\,\,$\Theta _{mean}$(GeV) \,\,\,\,&
\,\,\,\,$T_{e,0}$(GeV$^{5}$)\,\,\,\,
& \,\,\,\,$\Pi _{0}$(GeV$^{5}$)\,\,\,\, & \,\,\,$w_{DE,0}$ \,\,\,
\\
\hline \hline $0.21$ & $10$ & $10^{-8}$ & $10^{-98}$ & $-10^{-71}$ &
-1\\
\hline $0.18$ & $10$ & $3.5\!\times\!10^{-10}$ & $10^{-103}$ &
$-10^{-71}$ & -1
\\
\hline $0.13$ & $10$ & $10^{-13}$ & $10^{-118}$ & $-10^{-71}$ & -1
\\
\hline $0.07$ & $50$ & $10^{-9}$ & $10^{-104}$ & $-10^{-69}$ & -1
\\
\hline $0.05$ & $50$ & $10^{-12}$ & $10^{-116}$ & $-10^{-69}$ & -1
\\
\hline $0.01$ & $100$ & $10^{-10}$ & $10^{-109}$ & $-10^{-68}$ & -1
\\
\hline $0.001$ & $120$ & $10^{-10}$ & $10^{-109}$ & $-10^{-68}$ & -1
\\ \hline
\end{tabular}

\vspace{0.4cm}
{\bf{Table 1:}} Summary of results for various values of the parameters consistent with today acceleration

\end{table*}
\end{center}

\vspace{-1cm}

The auxiliary field $\psi$ which is the basic component of the dark energy can also be estimated during
the cosmic evolution ($\psi$ appears in equation
(\ref{systemb2})). Numerical results show that $\psi$ decreases during the evolution from $z=2$ to 0 with
typical values in the range
$ 2.5\times 10^{-83}$GeV$^{2}$  $<\psi<5\times 10^{-84}$GeV$^{2}$,
while $2 \gamma\, \rho$ also decreases and takes values in the region
$2.8\!\times\!10^{-83} $GeV$^{2}$
$<2\gamma\, \rho < 4.5\!\times\!10^{-84}$GeV$^{2}$. It is now apparent that the small outflow that
produces the values of dark pressure shown in
Table 1, is associated with values of $\psi$ comparable
with $\gamma\,\rho$ values, and consequently modify non trivially
the cosmic expansion and cosmic acceleration through equations
(\ref{systemb2}), (\ref{accel}).
Note that in the absence of outflow (in our case before the structure formation) the usual braneworld
cosmology holds with the well known solution $\mathcal{C}/a^{4}$ for $\psi$, which is the so called
dark radiation. Assuming a typical ansatz for the law that describes the non linear increase of the galactic core black hole mass density
from $z=8$ (approximate moment of the formation of first galaxies) to $z=2$ (moment that observations
suggest a known value for $\rho _{coreBH}$) it was possible to show that starting from a zero or from a small positive dark radiation, dark energy $\psi$ increases to the values described above (for $z\leqslant 2$) which are capable to drive cosmic acceleration.
Moreover, this positive radiation term $\mathcal{C}/a^{4}>0$ at $z=8$ is small enough to overpasses
well known problems of nucleosynthesis constraints \cite{ichicki}.

\begin{figure}[h!]
\centering
\begin{tabular}{cc}
\includegraphics*[width=220pt, height=150pt]{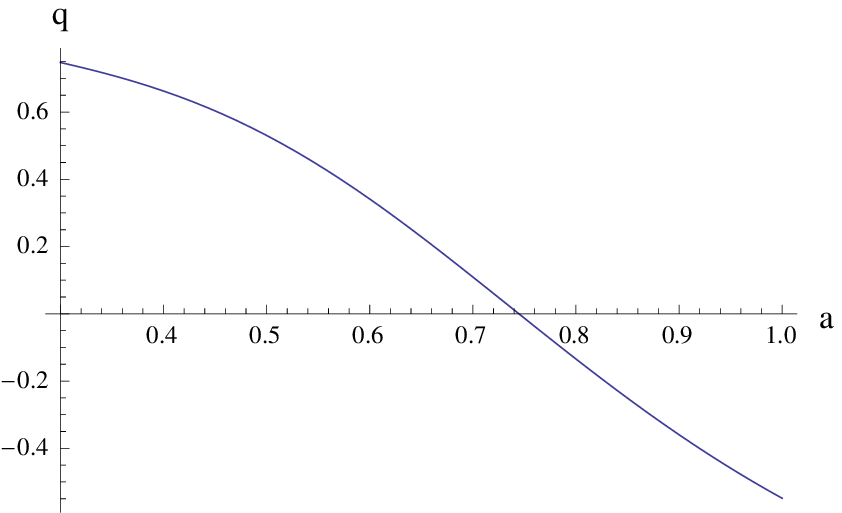}\\
\includegraphics*[width=220pt, height=150pt]{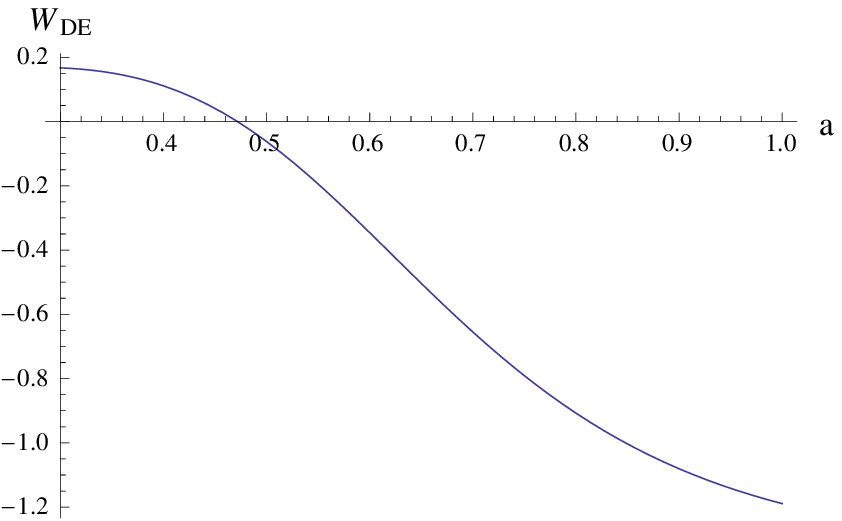}
\end{tabular}
 \caption{
A typical evolution of $q$ and $w_{DE}$ as a function of the scale
factor $a$}
\end{figure}


In order to confirm these numerical results, the analytical solution for the system of differential equations (\ref{systemb1})-(\ref{systemb3}) has been derived. The analytic solution that will be given corresponds to the case $T=0$. Indeed this approximation is valid since for all the interesting cosmological parameters (which make the scenario successful) the contribution of the terms containing $T$ is negligible compared to the other terms.
The solution is
\begin{eqnarray}
\label{analytic}
&&\!\!\!\!\!\!\psi(a)=\frac{H_{0}^2}{a^4}\Big{(}\Omega_{DE,0}-\frac{\beta
H_{0}^2\Omega_{m,0}^2}{4
\gamma^2}\Big{)}-\frac{\sqrt{\beta}\,\xi}{12}\frac{(\rho_{pls}^{cBH})^{\hat{\gamma}}}{M^3}
\rho_{cBH,0}\nn \\
&&\times\Big{\{}10^{\mu(1-\frac{1}{a})}\Big{[}\Big(\frac{\mu\ln{\!10}}{a}\Big)^{3}
\!-\Big(\frac{\mu\ln{\!10}}{a}\Big)^{2}
+\frac{2\mu\ln{\!10}}{a}-6\Big{]}\nn
\\&&\,\,\,\,
-\frac{1}{a^{4}}[(\mu\ln{\!10})^{3}-(\mu\ln{\!10})^{2}+2\mu\ln{\!10}-6]\nn\\
&&\,\,\,\,
+(\mu\ln{\!10})^{4}\frac{10^{\mu}}{a^{4}}\Big[Ei\Big(\!\!-\frac{\mu\ln{\!10}}{a}\Big)
\!-\!Ei(-\mu\ln{\!10})\Big]\Big{\}}\\
&&\!\!\!\!\rho(a)=\frac{H_{0}^{2}\Omega_{m,0}}{2\gamma a^{3}}\\
&&\!\!\!\!H^{2}(a)=\beta\rho^{2}+2\gamma\rho+\psi,
\end{eqnarray}
where $\rho_{cBH,0}=\rho_{coreBH}|_{z=0}$ and $Ei$ is the
exponential integral function. Based on this solution the results
shown in Table 1 are verified.

{\it{II. Numerical analysis with a certain number of black holes}}.
A different useful approach although less general, is to study the behavior
of our mechanism for the recent epoch $z\sim 0.$ If we are not interested to
investigate the early time evolution of the cosmic acceleration but just to
explore parameter combinations that provide values $w_{DE,0}\simeq -1$, it
is correct to set in the relevant differential equations
(\ref{systemc1}), (\ref{systemc2}) constant values for $T$, $\Pi$.
These values are estimated taking into account the present
values of astrophysical data (number of galaxies, number of black
holes etc.). The black hole mass density decreases due to
the cosmic expansion and increases due to matter accretion but
for the current time period of interest ($z<1$) the cosmic matter
density rate is three orders of magnitude larger than the black hole
density rate $d\rho _{BH}/dt<<d\rho /dt$ \cite{matteo}.
Therefore it is a fair approximation to assume for the recent cosmic evolution,
constant values for $T$, $\Pi$ in the differential equations (\ref{systemc1}),
(\ref{systemc2}).

\begin{center}
\begin{table*}
\begin{tabular}[t]{|c|c|c|c|c|c|}
\hline\hline \multicolumn{1}{||c|}{astrophysical observation} &
\multicolumn{1}{||c|}{ assumption} &
\multicolumn{1}{||c|}{assumption} & \multicolumn{1}{||c|}{ output} &
\multicolumn{1}{||c|}{output} & output \\ \hline\hline
$N_{BH}M_{BH}$ & $\hat{\gamma}$ & \,\,\,\,$\Theta _{mean}$(GeV)\,\,\,\,
&\,\,\,\, $T_{e,0}$(GeV$^{5}$)\,\,\,\, &\,\,\,\, $\Pi _{0}$(GeV$^{5}$)\,\,\,\, & \,\,\,$%
w_{DE,0}$\,\,\, \\ \hline $N_{coreBH}M_{BHcore}=10^{18}M_{\odot }$ &
$0.1$ & $10^{-10}$ & $10^{-105}$ & $-10^{-69}$ & -1
\\ \hline $N_{coreBH}M_{BHcore}=10^{18}M_{\odot }$ & $0.07$ &
$10^{-14}$ & $10^{-121}$ & $-10^{-69}$ & -1 \\
\hline
$N_{coreBH}M_{BHcore}=10^{15}M_{\odot }$ & $0.026$ & $10^{-8}$ & $%
10^{-101}$ & $-10^{-69}$ & -1 \\ \hline
$N_{haloBH}M_{haloBH}=10^{23}M_{\odot }$ & $0.2$ & $4.5\!\times\!
10^{-12}$ & $10^{-106}$ & $-10^{-69}$ & -1 \\
\hline
\end{tabular}

\vspace{0.4cm} {\bf{Table 2:}} Summary of results for $M=50\,TeV$ and for
various values of the parameters consistent with today acceleration

\end{table*}
\end{center}

\vspace{-0.8cm}
Based on the derived expressions (\ref{T0}), (\ref{P0}) for the cosmic
energy outflow $T_{e,0}$ and the associated pressure $\Pi _{0}$, we
will consider various cases for the black hole matter content in order to
evaluate the cosmic acceleration.

First we will consider as an extreme case a matter content with a large
amount of black holes in halos suggested in \cite{frampton}. In this case,
we assume a universe with $10^{11}$ halos and $10^{10}$ large black holes\
per halo. Further we set as a crude mean mass for a halo black hole a value
equal to $M_{haloBH}=10^{2}M_{\odot }$. Consequently, we estimate a total
mass in the form of halo black holes equal to $%
N_{haloBH}M_{haloBH}=10^{23}M_{\odot }$ (these numbers were taken from \cite%
{frampton}, however, note that the assumption appeared in \cite{frampton}
that all dark matter consists of black holes is not necessary or related to
the present paper). Galactic core black holes contribute much less in this case, i.e.
there are $10^{11}$ supermassive black holes each with a mean mass $%
10^{7}M_{\odot }$, i.e. $N_{coreBH}M_{BHcore}=10^{18}M_{\odot }.$
Therefore, in this extreme case all the cosmic acceleration comes
from halo black holes. We can get $w_{DE,0}\simeq -1$ for various combinations of the parameters, see Table 2.

It is more safe to assume that the mass density of galactic core black holes
is larger than the density of halo black holes. Taking $%
N_{coreBH}M_{coreBH}=10^{18}M_{\odot }$, again there are plenty of
numerical solutions of (\ref{systemc1}), (\ref{systemc2}) for
various parameters giving acceleration $w_{DE,0}\simeq -1$. Similarly, assuming a more conservative case where $%
N_{coreBH}M_{coreBH}=10^{15}M_{\odot }$ it is easy to find many
numerical solutions of equations (\ref{systemc1}), (\ref{systemc2}) resulting
to the required cosmic acceleration. Such representative results are shown in Table 2.
Results shown in Table 2 reveal that the various quantities are of the same order with the corresponding quantities of Table 1 as expected.

{\it{III. Astrophysical constraints}}. It is worth emphasizing that all these estimated values of energy
loss $T_{e,0}$ appeared in Tables 1 and 2 are small values that do
not cause any astrophysical inconsistency on galaxy evolution or
black hole dynamics. The rest of this section is devoted to the
explanation of the absence of any conflict with the known
observational characteristics of the galaxies and of their black
holes. Thus, two astrophysical constraints are considered for the case where the outflow
occurs at the centers of core black holes. The other case where galactic halo black holes
dominate the energy outflow will be considered later.

A first bound can be found demanding that the lifetime of a galactic
core black hole loosing energy according to our scenario is larger
than the typical lifetime of such black holes
$t_{coreBH}\sim 10^{10}$ years. An estimate of the
lowest possible lifetime in the worst case scenario is found dividing the rest black hole energy $M_{BH}$ of a typical black hole by its outflow, eq.(\ref{escapeplasma}), and its volume, therefore
\begin{equation}
t_{\text{lifetime}}=\frac{2M^3}{0.112 \Theta _{mean}^{4}}>t_{coreBH}.
\end{equation}
This bound is easily satisfied for all expected values of $T_{e,0}$,
for example for $M \sim 50 TeV$ and $\Theta _{mean}\sim
10^{-9}GeV$ the lifetime estimate is $10^{20}$ years!

A second bound can be obtained from the requirement that the current
total energy loss from all the volume of black hole be smaller than the energy gain from the accretion of
the black hole at the galactic core minus the energy ejected in
various frequencies.  In galaxies with AGNs and a super massive
black hole \cite{Agn1}, \cite{AGN2}, \cite{AGN3} the mass accretion
rate $\dot{M}_{BH}$ is expected to be a fraction of the measured
luminosity $L$. The ratio $L/\dot{M}_{BH}$ defines the conversion
efficiency of gravitational energy into radiation and varies during
the evolution of the accretion disc within the range
$10^{-3}-10^{-1}$. On the other hand, the measured luminosity has
been observed to be always a fraction (called efficiency and ranging
from somewhat below 0.01 for the low luminosity AGNs to 0.1 for the
large luminosity AGNs - strong accretors) of the Eddington
luminosity $L_{E}\sim\frac{M_{BH}}{10^{8}M_{\odot }}10^{46}$erg
sec$^{-1}$. As a result, in any case for the purpose of estimating
this second constraint, the net gain of energy rate $\dot{M}_{BH}$
can be assumed to be around $L_{E}$ and therefore the second bound becomes
\begin{equation}\label{secondbound}
\Delta\dot{\rho}^{BH}_{pls} V_{BH}=0.112\frac{\Theta _{mean}^{4}}{2M^3}M_{BH}\ll L_{E}.
\end{equation}
A black hole with mass $M_{BH}\sim 10^{7}M_{\odot }$,
losing energy with average temperature of the effective plasma $\Theta _{mean}\sim10^{-9} GeV$ and $M = 50 TeV$,
is associated with an energy loss rate equal to
$10^{34}$ erg sec$^{-1}$. Such losses are orders of magnitude smaller numbers compared to accretion rates.
Therefore, they cannot alter significantly the
black hole mass and make impossible the violation of any measured
relations between central galactic back hole mass and galactic halo
mass or of the observed expression of black hole density as a
function of redshift (eq. (\ref{rhoofz}) that was used in the present section).

Next, let's study the constraints regarding the extreme case where
galactic halo black holes \cite{binary}, \cite{binary2} dominate the
energy outflow $T_{e,0}$. The first bound demands that the time
duration required for a black hole of mass $M_{BH}$ to lose all its
rest energy be larger than the maximum lifetime of a typical
galactic halo black hole $t_{haloBH}\sim 10^{10}$ years. The first bound can be
easily met; for example, for a halo black hole (see Table 2) with $\Theta _{mean}\sim 10^{-12} GeV$
the estimated lifetime is $10^{31}$years!

Finally, for halo black holes the second bound can be studied
demanding the net energy gain due to accretion of mass minus the
radiated energy be larger than the energy loss to extra dimensions.
Now, we have to distinguish two types of galactic halo black holes.
Black holes that are part of a binary system have usually an
efficiency $L/L_{E}$ from 0.01 to 1, while the conversion efficiency
is around 0.01 to 0.1. Thus the net energy rate gain is expected to be close to Eddington
accretion, which in this case is $L_{E}\sim 10^{40}$erg sec$^{-1}$.
For halo black holes with mass $M_{haloBH}\sim 10^{2}M_{\odot }$ and with $\Theta _{mean}\sim 10^{-12} GeV$ the loss rate is equal only to $10^{17}$ erg sec$^{-1}$. On the other hand, galactic halo black holes that do not belong to a binary system
cannot be observed since they do not accrete matter and there is no
accretion disk to radiate. Therefore, in this case it is not
possible to know their properties and apply the second bound.

It is worth mentioning that our cosmology solves also the problem of the age of the universe. This is known to be true for braneworld models with non zero $T, \Pi$ due to the
produced cosmic acceleration \cite{bounds}.

\section{Discussion and Conclusions}

This work proposes that the recent cosmic acceleration happens as a result of the negative five-dimensional pressures produced in the galactic black holes of a brane cosmological RS setup. It was proved that the total contribution $\Pi$
from all the dark pressures from all the centers of galaxies suffice
to provide the measured cosmic acceleration. An exciting outcome is that the recent passage from the deceleration to the acceleration era happens due to the recent increase of the galactic core black hole mass density. Based on the derived expressions we have shown that
it is easy to get the expected negative values of cosmic
acceleration $w_{DE,0}\simeq -1$ and dark energy $\Omega_{DE,0}=0.7$ \emph{even for conservative values of
all the relevant parameters}, i.e. for small values of the mean temperature $\Theta _{mean}$ in the interior of the black holes, for small values of galactic core black holes masses and for values of the five-dimensional Planck mass $M$ several decades of TeVs. In our mechanism, outflow is associated unavoidably with a large dark
pressure. The magnitude of the produced dark pressure is
connected with that of dark radiation through the equation of state
of the quantum fluid in the interior of a black hole. Qualitatively one can
immediately check the efficiency of producing the observed cosmic
acceleration estimating the required amount of dark pressure shown
in equations (\ref{q0}), (\ref{current po}) or (\ref{current po2}).
This value of dark pressure can naturally be realized in our case based on the known astrophysical data and the assumed temperature of the effective fluid. Of course, in order cosmic acceleration to be
proven, the system of differential equations (\ref{systemc1}),
(\ref{systemc2}) has to be solved, and indeed it is seen that the
energy exchange along with the dark pressure give an order one
effect on cosmological scales.

The proposed mechanism has several advantages: i) it is independent
of the bulk matter and consequently retains predictability, ii) the
associated values of $\psi$ in the Hubble evolution
(\ref{systemb2}) originate from the brane black hole astrophysical
phenomenon of energy outflow $T$ and its associated pressure $\Pi $
along the fifth dimension, and not from the motion or the position
of the brane in the bulk, thus again retaining predictability, iii)
the mechanism is \textquotedblleft on" at present times and
\textquotedblleft off" at the early stages of the cosmic evolution
explaining naturally coincidence problem, iv) it relates the amount
of the produced acceleration with the present matter content, and v)
it produces easily cosmic acceleration for sensible values of the relevant parameters.

However, the most interesting and worth mentioning finding is the
fact that sensible and safe values of outflow, as those appeared
in Table 2, suffice to result to the observed cosmic acceleration.
A first reason behind this outcome is the large number
of galaxies in the universe. Another reason is that the additive effect of all outflows and associated much larger dark pressures from each galactic center result to a non negligible kinetic effect, i.e. acceleration due to the geometry of the setup. Finally, the
dark pressure drives towards acceleration from earlier times of the cosmic evolution and not just today. Since energy density of the
galactic core black holes increases as redshift decreases at recent
times ($z<2$), dark pressure becomes stronger driving the
passage to the acceleration era.

In summary, the novelties of the present article are: 1) correction of a wide spread mistake among the brane cosmologists that outflow is associated with a positive and not negative dark pressure, 2) the presentation of
a new astrophysical origin mechanism of brane-bulk energy exchange, 3) new braneworld solutions describing the evolution of brane equations with non zero $T_{05}$, $ T_{55}$, 4) new gravitational collapse solution on a brane with non zero $T_{05}$, $T_{55}$, 5) numerical results estimating the produced
cosmic acceleration, and 6) correlation of the measured acceleration to the recent rise of the galactic core black hole cosmic energy density.

The calculations of the scenario could have failed for various
reasons: if a very high temperature was needed in the interior of
the black hole, or if a small fundamental Planck scale or a large
$T^{0}_{5}$ was needed conflicting of course with galactic dynamics,
or if for the given variation of the cosmic black hole density as a
function of redshift the cosmic evolution failed to posses a long
deceleration era accompanied by a recent acceleration one. However,
the concrete and conservative numerical values used lead the
scenario to success.


One interesting point worth to be raised is that the proposed
mechanism could work together with various studies that suggest solutions to the cosmological problem. For example, there are recent holographic ideas capable to explain why the cosmological constant should be almost zero \cite{thomas}, \cite{hsu}. However, it appears to have a difficulty to explain naturally the cosmic equation of state. Therefore, the present work together with all type of holographic explanations can provide a complete solution to the general problem of the cosmological constant value.

\section{Acknowledgements}

We would like to thank V. Charmandaris, T. Harko, E. Kiritsis and I.
Papadakis for useful discussions and comments.

\end{document}